\pdfoutput=1
\documentclass[prb,twocolumn,showpacs,aps]{revtex4}
\usepackage{graphicx,graphics,color,epsfig}
\usepackage{bm}
\usepackage{amsmath}
\usepackage{amssymb}
\usepackage{epstopdf}

\makeatletter

\newcommand{\Rmnum}[1]{\expandafter\@slowromancap\romannumeral #1@}
\newcommand*{\rom}[1]{\expandafter\@slowromancap\romannumeral #1@}
\makeatother

\usepackage{color}

\begin{document}

\title{Possible spin excitation structure in monolayer FeSe grown on SrTiO$_{3}$}

\author{Yi Gao,$^{1}$ Yan Yu,$^{1}$ Tao Zhou,$^{2}$ Huaixiang Huang,$^{3}$ and Qiang-Hua Wang $^{4,5}$}
\affiliation{$^{1}$Department of Physics, Nanjing Normal University, Nanjing, 210023, China\\
$^{2}$College of Science, Nanjing University of Aeronautics and Astronautics, Nanjing, 210016, China\\
$^{3}$Department of Physics, Shanghai University, Shanghai, 200444, China\\
$^{4}$National Laboratory of Solid State Microstructures $\&$ School of Physics, Nanjing
University, Nanjing, 210093, China\\
$^{5}$Collaborative Innovation Center of Advanced Microstructures, Nanjing 210093, China}

\begin{abstract}
Based on recent high-resolution angle-resolved photoemission spectroscopy measurement in monolayer FeSe grown on SrTiO$_{3}$, we constructed a tight-binding model and proposed a superconducting (SC) pairing function which can well fit the observed band structure and SC gap anisotropy. Then we investigated the spin excitation spectra in order to determine the possible sign structure of the SC order parameter. We found that a resonance-like spin excitation may occur if the SC order parameter changes sign along the Fermi surfaces. However, this resonance is located at different locations in momentum space compared to other FeSe-based superconductors, suggesting that the Fermi surface shape and pairing symmetry in monolayer FeSe grown on SrTiO$_{3}$ may be different from other FeSe-based superconductors.
\end{abstract}


\maketitle
\section{introduction}
In most iron pnictide superconductors, there exist hole and electron pockets in the Brillouin zone (BZ) center and corner, respectively and it is commonly accepted that the spin fluctuation induced by the nesting between the hole and electron pockets leads to the unconventional superconductivity and high transition temperature T$_c$. In this case, the superconducting (SC) pairing symmetry is $s_\pm$-wave, where the sign of the SC order parameter on the hole pockets is opposite to that on the electron ones.\cite{mazin} At early stages it was believed that, if either the hole or electron pocket disappears, then the nesting condition between the hole and electron pockets will be broken and superconductivity will be suppressed, as in the hole-overdoped Ba$_{1-x}$K$_x$Fe$_{2}$As$_{2}$ \cite{ding1} and electron-overdoped BaFe$_{2-x}$Co$_{x}$As$_{2}$.\cite{Takahashi} In this regard, the recent findings of high-T$_c$ superconductivity in a family of FeSe-based iron pnictide superconductors, such as Li$_{1-x}$Fe$_x$OHFe$_{1-y}$Se,\cite{chenxh,Johrendt,zhaozx} Li$_x$(NH$_2$)$_y$(NH$_3$)$_{1-y}$Fe$_2$Se$_2$ \cite{clarke} and A$_x$Fe$_{2-y}$Se$_2$ (A=Rb, Cs, K),\cite{chenxl,chenxh2,conder} as well as monolayer FeSe grown on SrTiO$_{3}$,\cite{xueqk1} are quite surprising and have attracted much attention among the community, since in these materials, there are no hole pockets, but only electron ones in the BZ.\cite{zhouxj1,zhouxj2,fengdl1,fengdl2,shenzx1,zhouxj3,fengdl4,ding,zhouxj4,fengdl5,fengdl6}

At present, the SC pairing symmetry in the FeSe-based superconductors is hotly debated. Theoretical investigations suggest that, in the absence of the hole pockets, the nesting is now between the electron ones and in this case, a nodeless $d$-wave pairing symmetry is the leading candidate,\cite{aoki,scalapino,leedh} which, when folded into the 2Fe/cell BZ, will lead to gap nodes or extreme minima in the vicinity of the BZ boundaries.\cite{mazin2} Experimentally, most angle-resolved photoemission spectroscopy (ARPES) measurements found that the SC gap magnitude is nearly isotropic along the Fermi surfaces, with no apparent nodes or extreme minima.\cite{zhouxj1,shenzx1,zhouxj3,fengdl4,zhouxj4,fengdl5,fengdl6} Inelastic neutron scattering (INS) experiments observed a spin resonance in Li$_{1-x}$Fe$_x$ODFe$_{1-y}$Se,\cite{Boothroyd,zhaoj1} Li$_x$(ND$_2$)$_y$(ND$_3$)$_{1-y}$Fe$_2$Se$_2$ \cite{Boothroyd1} and A$_x$Fe$_{2-y}$Se$_2$ (A=Rb, Cs, K),\cite{Inosov1,Inosov2,Boothroyd2,Inosov3,zhaoj} which was interpreted as the signature of the sign change of the SC order parameter. However, scanning tunneling microscopy (STM) experiments performed in Li$_{1-x}$Fe$_x$OHFe$_{1-y}$Se \cite{fengdl7} and monolayer FeSe grown on SrTiO$_{3}$ \cite{fengdl3} ruled out any sign change of the SC order parameter along the Fermi surfaces.

Among the above mentioned FeSe-based superconductors, monolayer FeSe grown on SrTiO$_{3}$ is of particular interest. T$_c$ in this material is above 50K,\cite{xueqk1} and can even reach up to 100K, \cite{jiajf} the highest among all iron pnictide superconductors. It has the simplest structure where the Fe atoms form a strictly two-dimensional (2D) lattice, with the Se atoms sitting below and above the Fe plane. In contrast to its simple structure, the SC mechanism is complicated and the high T$_c$ may be resulted from a combination of the electron-electron interaction within the monolayer FeSe and the electron-phonon interaction between FeSe and the SrTiO$_{3}$ substrate,\cite{shenzx1,leedh2} where the interface effect may play a vital role in enhancing T$_c$. Recently its Fermi surfaces have been precisely mapped out by high-resolution ARPES and clear gap anisotropy has been observed, putting strong constraint on the possible pairing symmetry.\cite{shenzx2} Furthermore, since ARPES measures only the magnitude of the SC gap, but not its phase, a phase-sensitive measurement is thus needed to determine the exact pairing symmetry. Therefore in this work, we investigate the spin excitation spectra in monolayer FeSe grown on SrTiO$_{3}$, which can be measured by INS, to determine the possible sign structure of the SC order parameter.

\section{method}
\begin{figure}
\includegraphics[width=1\linewidth]{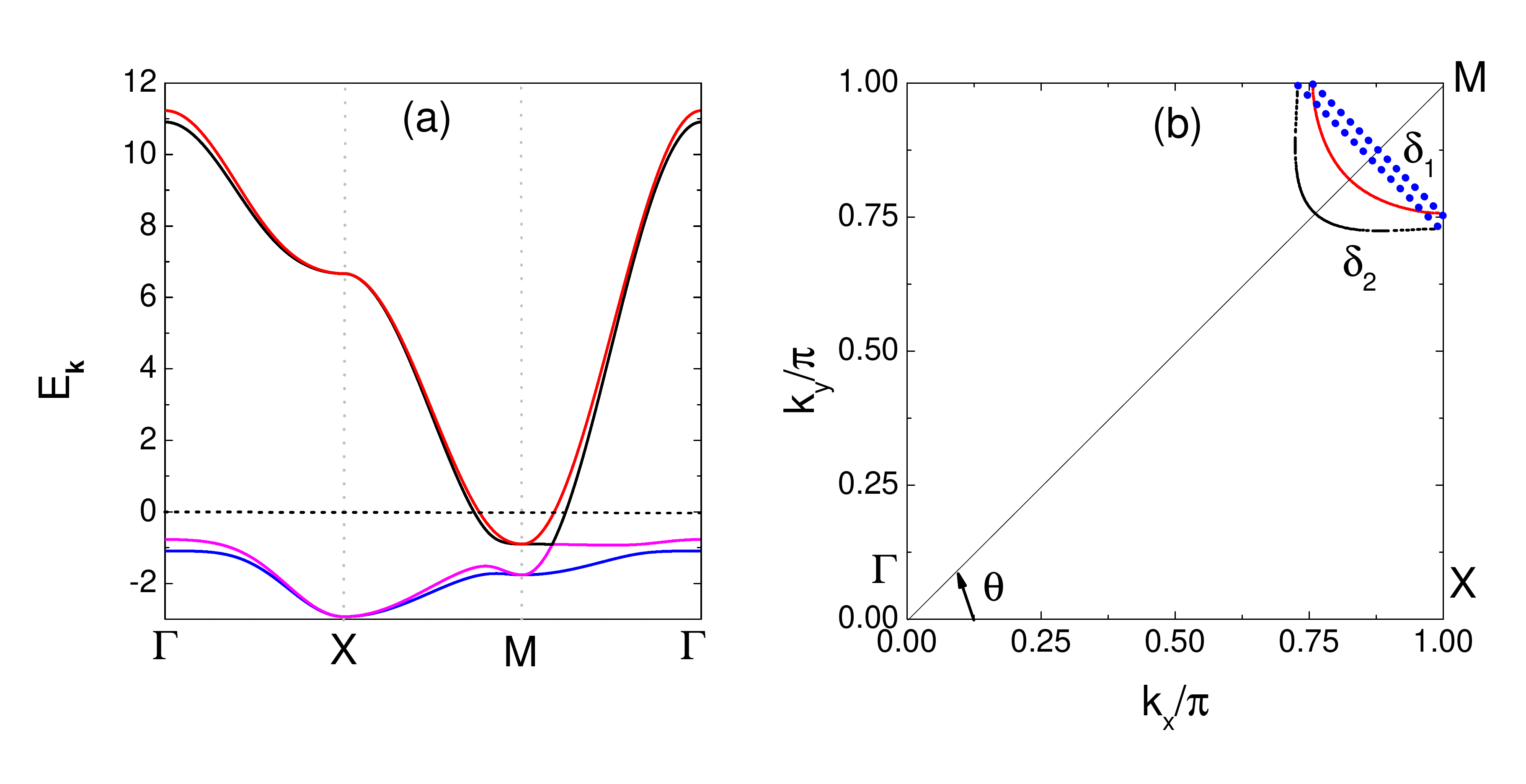}
 \caption{\label{band} (color online) (a) Calculated band structure along the high-symmetry directions in the 2Fe/cell BZ. The energy is defined with respect to the Fermi level (the black dotted line). (b) The Fermi surfaces in the first quadrant of the 2Fe/cell BZ (black and red), while the blue dotted lines show schematically the Fermi surfaces adopted in Refs. \onlinecite{scalapino}, \onlinecite{Boothroyd} and \onlinecite{Inosov2}.}
\end{figure}

We adopt a 2D tight-binding model proposed in Ref. \onlinecite{gaoy_fese}, where each unit cell contains two inequivalent sublattices $A$ and $B$. The coordinate of the sublattice $A$ in the unit cell $(i,j)$ is $\mathbf{R}_{ij}=(i,j)$ while that for the sublattice $B$ is $\mathbf{R}_{ij}+\mathbf{d}$, with $\mathbf{d}$ being $(0.5,0.5)$. The Hamiltonian can be written as $H=\sum_{\mathbf{k}}\psi_{\mathbf{k}}^{\dag}A_{\mathbf{k}}\psi_{\mathbf{k}}$, where
\begin{eqnarray}
\label{h}
\psi_{\mathbf{k}}^{\dag}&=&(c_{\mathbf{k}A1\uparrow}^{\dag},c_{\mathbf{k}A2\uparrow}^{\dag},c_{\mathbf{k}B1\uparrow}^{\dag},c_{\mathbf{k}B2\uparrow}^{\dag},\nonumber\\
&&c_{-\mathbf{k}A1\downarrow},c_{-\mathbf{k}A2\downarrow},c_{-\mathbf{k}B1\downarrow},c_{-\mathbf{k}B2\downarrow}),\nonumber\\
A_{\mathbf{k}}&=&\begin{pmatrix}
M_{\mathbf{k}}&D_{\mathbf{k}}\\D_{\mathbf{k}}^{\dag}&-M_{-\mathbf{k}}^{T}
\end{pmatrix},\nonumber\\
M_{\mathbf{k}}&=&\begin{pmatrix}
\epsilon_{A,\mathbf{k}}&\epsilon_{xy,\mathbf{k}}&\epsilon_{T1,\mathbf{k}}&0\\
\epsilon_{xy,\mathbf{k}}&\epsilon_{A,\mathbf{k}}&0&\epsilon_{T2,\mathbf{k}}\\
\epsilon_{T1,\mathbf{k}}&0&\epsilon_{B,\mathbf{k}}&\epsilon_{xy,\mathbf{k}}\\
0&\epsilon_{T2,\mathbf{k}}&\epsilon_{xy,\mathbf{k}}&\epsilon_{B,\mathbf{k}}
\end{pmatrix}.
\end{eqnarray}
Here $c_{\mathbf{k}A1\uparrow}^{\dag}/c_{\mathbf{k}A2\uparrow}^{\dag}$ creates a spin up electron with momentum $\mathbf{k}$ and on the $d_{xz}/d_{yz}$ orbital of the sublattice $A$. $\epsilon_{A,\mathbf{k}}=-2(t_{3}\cos k_{x}+t_{4}\cos k_{y})-\mu$, $\epsilon_{B,\mathbf{k}}=-2(t_{3}\cos k_{y}+t_{4}\cos k_{x})-\mu$, $\epsilon_{xy,\mathbf{k}}=-2t_{5}(\cos k_{x}+\cos k_{y})$, $\epsilon_{T1,\mathbf{k}}=-2t_{1}\cos[(k_{x}+k_{y})/2]-2t_{2}\cos[(k_{x}-k_{y})/2)]$ and $\epsilon_{T2,\mathbf{k}}=-2t_{1}\cos[(k_{x}-k_{y})/2]-2t_{2}\cos[(k_{x}+k_{y})/2)]$. In addition, $M_{\mathbf{k}}$ and $D_{\mathbf{k}}$ are the tight-binding and pairing parts of the system, respectively. Throughout this work, the momentum $\mathbf{k}$ is defined in the 2Fe/cell BZ and the energies are in units of 0.1eV. In the following we set $t_{1-5}=1.6,1.4,0.4,-2,0.04$ and $\mu=-1.87$ to fit the band structure measured by ARPES. Under this set of parameters, the average electron number is $n\approx2.1$ (the system is about $10\%$ electron doped). Fig. \ref{band} shows the calculated band structure and Fermi surfaces. The $\Gamma$ hole pockets sink below the Fermi level while two electron pockets $\delta_{1}$ and $\delta_{2}$ exist around $M$ with their sizes similar to the ARPES data ($k_{F}/\pi\approx0.25$). A slight ellipticity in the electron pockets $\delta_{1}$ and $\delta_{2}$ is induced by setting $t_{1}\neq t_{2}$. Both the electron number and the Fermi surface topology are consistent with ARPES.\cite{zhouxj1,shenzx2} The reason we fit our tight-binding model to ARPES instead of the LDA band structure is that the bottom of the $M$ electron band in the LDA-calculated band structure is $\sim$500meV below the Fermi level, almost an order of magnitude deeper than the ARPES results.\cite{sadovskii} Furthermore, if the correlation effect can account for the shallow $M$ electron band, a hole pocket will show up around $\Gamma$.\cite{sadovskii2} In contrast, in our model, the top of the $\Gamma$ hole band and the bottom of the $M$ electron band are both located $\sim$100meV below the Fermi level, agreeing with ARPES much better than the LDA-calculated band structure.

The band structure as well as the pairing function in the band basis can be obtained through a unitary transform $Q_{\mathbf{k}}$ as
\begin{eqnarray}
\label{unitary}
Q_{\mathbf{k}}^{\dag}M_{\mathbf{k}}Q_{\mathbf{k}}&=&\begin{pmatrix}
E_{\mathbf{k}}^{1}&0&0&0\\
0&E_{\mathbf{k}}^{2}&0&0\\
0&0&E_{\mathbf{k}}^{3}&0\\
0&0&0&E_{\mathbf{k}}^{4}
\end{pmatrix},\\
\Delta_{\mathbf{k}}&=&Q_{\mathbf{k}}^{\dag}D_{\mathbf{k}}Q_{-\mathbf{k}}^{*}.
\end{eqnarray}
Here $E_{\mathbf{k}}^{1},E_{\mathbf{k}}^{2}$ are the energies of the two hole bands which sink below the Fermi level and $E_{\mathbf{k}}^{3},E_{\mathbf{k}}^{4}$ are those of the two electron bands ($E_{\mathbf{k}}^{1}<E_{\mathbf{k}}^{2}<E_{\mathbf{k}}^{3}<E_{\mathbf{k}}^{4}$). The diagonal components in $\Delta_{\mathbf{k}}$ represent the pairing function on each band while the off-diagonal components signify the inter-band pairing. Since we are interested in the spin excitation spectra, we calculate the multiorbital dynamical spin susceptibility as \cite{kuroki}
\begin{widetext}
\begin{equation}
\label{spin_susceptibility}
\chi^{r\alpha,s\beta,t\gamma,u\delta}(\mathbf{q},i\omega_{n})=\frac{1}{N}\int_{0}^{\beta}d\tau e^{i\omega_{n}\tau}\langle T_{\tau}S^{+r\alpha,s\beta}_{\mathbf{q}}(\tau)
S^{-t\gamma,u\delta}_{-\mathbf{q}}(0)\rangle,
\end{equation}
\end{widetext}
with $S^{+r\alpha,s\beta}_{\mathbf{q}}=\sum_{\mathbf{k}}c^{\dag}_{\mathbf{k}r\alpha\uparrow}c_{\mathbf{k}+\mathbf{q}s\beta\downarrow}$ and the experimentally measured spin susceptibility is
\begin{eqnarray}
Im\chi(\mathbf{q},\omega+i\eta)&=&\sum_{r,t}\sum_{\alpha,\gamma}Im\chi^{r\alpha,r\alpha,t\gamma,t\gamma}(\mathbf{q},\omega+i\eta).\nonumber\\
\end{eqnarray}
Within the random-phase approximation (RPA) and writing the spin susceptibility in a matrix form, we have
\begin{eqnarray}
\label{rpa}
\hat{\chi}(\mathbf{q},\omega+i\eta)&=&\hat{\chi}^{0}(\mathbf{q},\omega+i\eta)[\hat{I}-\hat{U}_{s}\hat{\chi}^{0}(\mathbf{q},\omega+i\eta)]^{-1},\nonumber\\
\end{eqnarray}
with the bare spin susceptibility being
\begin{widetext}
\begin{eqnarray}
\label{x0}
\chi^{0r\alpha,s\beta,t\gamma,u\delta}(\mathbf{q},\omega+i\eta)&=&\frac{1}{4N}\sum_{o,p=1}^{4}\sum_{\mathbf{k}}Q_{\mathbf{k}}^{m_1o}Q_{\mathbf{k}}^{m_4o}Q_{\mathbf{k}+\mathbf{q}}^{m_2p}Q_{\mathbf{k}+\mathbf{q}}^{m_3p}
C^{op}(\mathbf{k},\mathbf{q})
(\frac{1}{\omega+i\eta+\xi_{\mathbf{k}}^{o}+\xi_{\mathbf{k}+\mathbf{q}}^{p}}-\frac{1}{\omega+i\eta-\xi_{\mathbf{k}}^{o}-\xi_{\mathbf{k}+\mathbf{q}}^{p}}),\nonumber\\
\end{eqnarray}
\end{widetext}
at $T=0$. Here $m_1=2(r-1)+\alpha$, $m_2=2(s-1)+\beta$, $m_3=2(t-1)+\gamma$, $m_4=2(u-1)+\delta$ and $C^{op}(\mathbf{k},\mathbf{q})=1-\frac{E_{\mathbf{k}}^{o}E_{\mathbf{k}+\mathbf{q}}^{p}+\Delta_{\mathbf{k}}^{o}\Delta_{\mathbf{k}+\mathbf{q}}^{p}}{\xi_{\mathbf{k}}^{o}\xi_{\mathbf{k}+\mathbf{q}}^{p}}$ is the usual coherence factor of the spin susceptibility. $\Delta_{\mathbf{k}}^{o}$ is the $o$th diagonal element of $\Delta_{\mathbf{k}}$ and $\xi_{\mathbf{k}}^{o}=\sqrt{E_{\mathbf{k}}^{o2}+\Delta_{\mathbf{k}}^{o2}}$. In deriving Eq. (\ref{x0}) we have neglected the off-diagonal components of $\Delta_{\mathbf{k}}$ and assumed that the orbital-band matrix element for the anomalous Green's functions is the same as that for the normal Green's functions, as done in Ref. \onlinecite{scalapino}.  We have verified that the above assumptions do not alter the spin susceptibility qualitatively, which is also found in Refs. \onlinecite{scalapino} and \onlinecite{siqm}. The nonzero elements of the interaction matrix $\hat{U}_s$ are $U_{s}^{r\alpha,r\beta,r\gamma,r\delta}=U$ for $\alpha=\beta=\gamma=\delta$, $U'$ for $\alpha=\delta\neq\beta=\gamma$, $J_H$ for $\alpha=\beta\neq\gamma=\delta$ and $J'$ for $\alpha=\gamma\neq\beta=\delta$.
In addition we have taken $J'=J_H=U/4$ and $U'=U-2J_H$.
From Eq. (\ref{x0}) we have, for $\eta=0^{+}$ and $\omega\geq0$,
\begin{widetext}
\begin{eqnarray}
\label{imx0}
Im\chi^{0r\alpha,s\beta,t\gamma,u\delta}(\mathbf{q},\omega+i\eta)&=&\frac{\pi}{4N}\sum_{o,p=1}^{4}\sum_{\mathbf{k}}Q_{\mathbf{k}}^{m_1o}Q_{\mathbf{k}}^{m_4o}Q_{\mathbf{k}+\mathbf{q}}^{m_2p}Q_{\mathbf{k}+\mathbf{q}}^{m_3p}
C^{op}(\mathbf{k},\mathbf{q})
\delta(\omega-\xi_{\mathbf{k}}^{o}-\xi_{\mathbf{k}+\mathbf{q}}^{p}),
\end{eqnarray}
\end{widetext}
and $Re\chi^{0}(\mathbf{q},\omega+i\eta)$ is related to $Im\chi^{0}(\mathbf{q},\omega+i\eta)$ by the Kramers-Kronig (KK) relations. In fully gapped superconductors,
for a fixed $\mathbf{q}$, due to the $\delta$ function in Eq. (\ref{imx0}), $Im\chi^{0}(\mathbf{q},\omega+i\eta)$ is zero when $\omega<E_{th}$, where $E_{th}$ is a threshold energy determined by the minimal value of $\xi_{\mathbf{k}}^{o}+\xi_{\mathbf{k}+\mathbf{q}}^{p}$. Meanwhile, if $C^{op}(\mathbf{k},\mathbf{q})\neq0$, then at $\omega=E_{th}$, there will exist a step discontinuity in $Im\chi^{0}(\mathbf{q},\omega+i\eta)$ which will lead to a logarithmic divergence in $Re\chi^{0}(\mathbf{q},\omega+i\eta)$. In this case, $Im\chi(\mathbf{q},\omega+i\eta)$ will diverge at $\omega=\omega_{res}$ (for arbitrary $U\neq0$), where $\omega_{res}\leq E_{th}$ and this signifies the formation of a spin resonance [at $\omega_{res}$, $Im\hat{\chi}^{0}=0$ and $det[\hat{I}-\hat{U}_sRe\hat{\chi}^{0}]=0$, leading to $Im\chi=\infty$ in Eq. (\ref{rpa})]. On the contrary, if $C^{op}(\mathbf{k},\mathbf{q})=0$, then the step discontinuity and the logarithmic divergence will disappear in $Im\chi^{0}$ and $Re\chi^{0}$, respectively, leading to the disappearance of any spin resonance for $\omega\leq E_{th}$.

For small but finite $\eta$, the step discontinuity in $Im\chi^{0}$ will become step-like, but nonvanishing and continuous across $E_{th}$, therefore the logarithmic divergence in $Re\chi^{0}$ will acquire a finite height. In this case, $det[\hat{I}-\hat{U}_sRe\hat{\chi}^{0}]=0$ can only be met with a sufficiently large $U$ and a peak with finite height instead of a divergence will show up in $Im\chi$, which also signifies a spin resonance.

In monolayer FeSe grown on SrTiO$_{3}$, there is no signature of static magnetic order when T$_c$ is the highest, therefore for arbitrary $\mathbf{q}$ in the normal state, we should have $det[\hat{I}-\hat{U}_sRe\hat{\chi}^{0}]\neq0$, which determines the upper bound of $U$ to be $6.8$ in our model [see Fig. \ref{gap}(a)]. On the other hand, since it may be close to a magnetic instability,\cite{zhouxj2,fengdl1,gong} therefore we set $U=6$ in the following. Here we need to point out that the value of $U$ in the present work ($\sim$0.6eV) seems to be smaller than those adopted in Refs. \onlinecite{aoki} and \onlinecite{scalapino} ($\sim$1eV). The reason is that, our band structure is a fit to the ARPES-measured one, which can be viewed as the LDA band structure renormalized by the correlation effect. We notice that the band width in Refs. \onlinecite{aoki} and \onlinecite{scalapino} is $\sim$4eV, while in our model it is $\sim$1.5eV, correspondingly a smaller $U$ in our model can lead to the magnetic instability.

\section{results and discussion}
\begin{figure}
\includegraphics[width=1\linewidth]{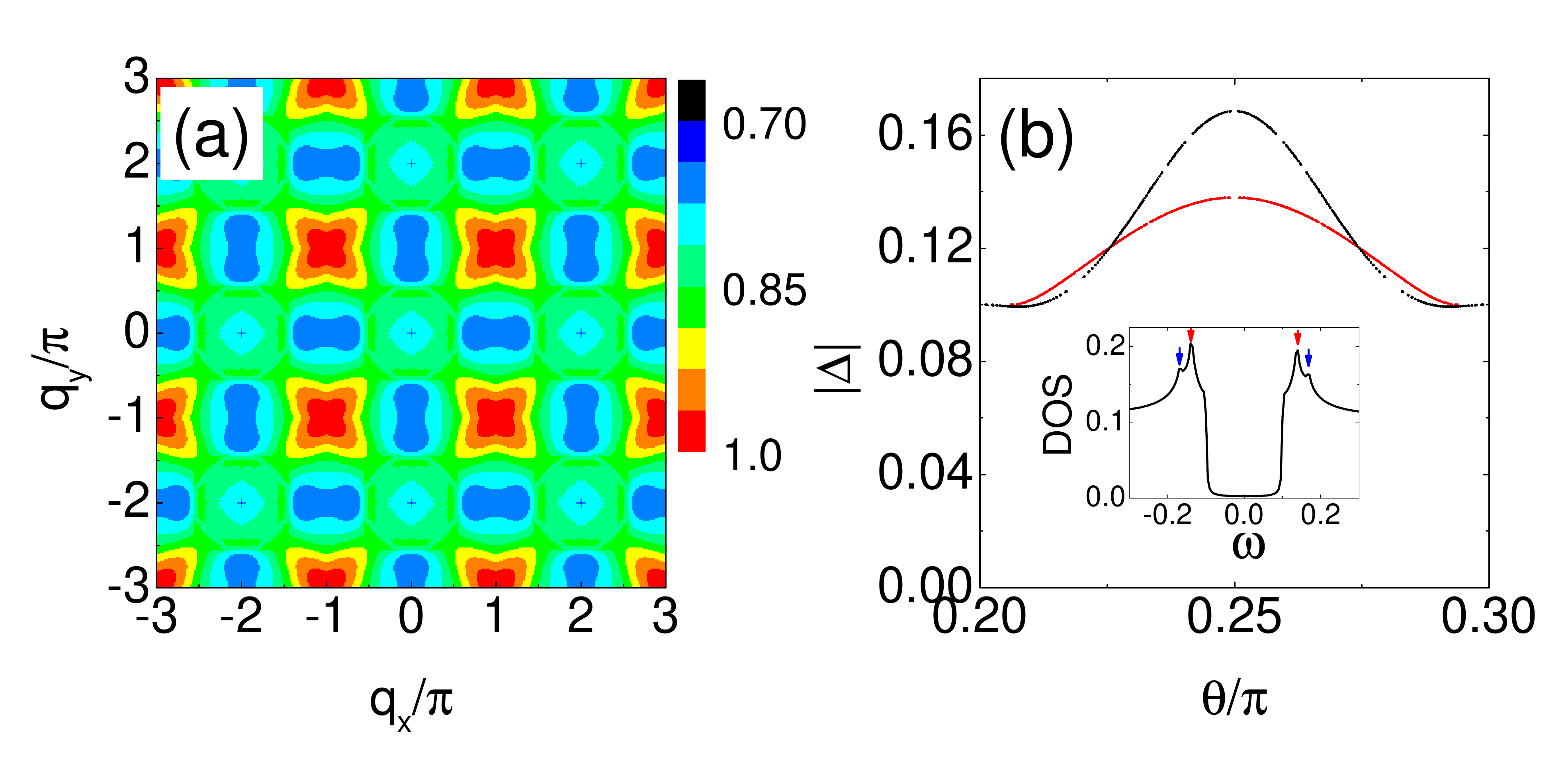}
 \caption{\label{gap} (color online) (a) The maximal eigenvalue of $\hat{U}_sRe\hat{\chi}^{0}(\mathbf{q},0+i\eta)$ in the normal state [$D_{\mathbf{k}}=0$ in Eq. (\ref{h})], by setting $U=6.8$. (b) The magnitude of the SC gap $|\Delta|$ along $\delta_{1}$ (red) and $\delta_{2}$ (black), defined as the minimal positive eigenvalue of $A_{\mathbf{k}}$ in Eq. (\ref{h}) along the Fermi surfaces. $\theta$ is defined in Fig. \ref{band}(b). The inset in (b) shows the DOS.}
\end{figure}

\begin{figure}
\includegraphics[width=1\linewidth]{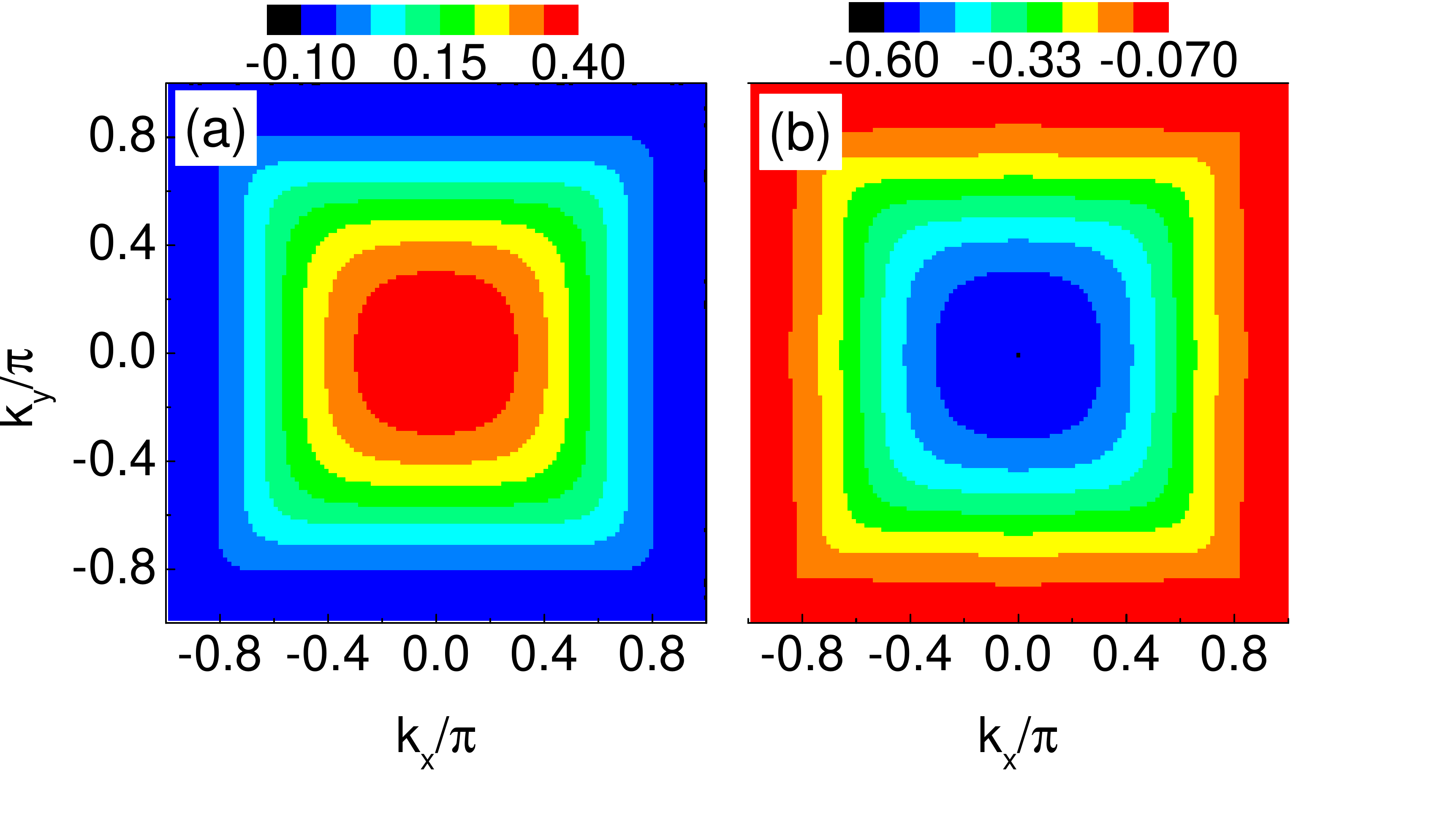}
 \caption{\label{gap_bz} (color online) The pairing function in the band basis. (a) $\Delta_{\mathbf{k}}^{1}$. (b) $\Delta_{\mathbf{k}}^{3}$. $\Delta_{\mathbf{k}}^{2}$ ($\Delta_{\mathbf{k}}^{4}$) is qualitatively the same as $\Delta_{\mathbf{k}}^{1}$ ($\Delta_{\mathbf{k}}^{3}$), but with some minor quantitative difference.}
\end{figure}

Having elucidated the origin of the spin resonance in the spin excitation spectrum, we then apply the above argument to our model of monolayer FeSe. In our previous work \cite{gaoy_fese} we have proposed a pairing function $D_{\mathbf{k}}$ that can well reproduce the anisotropy of the SC gap observed by ARPES,\cite{shenzx2} which can be expressed as
\begin{eqnarray}
\label{dk}
D_{\mathbf{k}}&=&\begin{pmatrix}
d_{0}&0&d_{1\mathbf{k}}&0\\
0&d_{0}&0&d_{1\mathbf{k}}\\
d_{1\mathbf{k}}&0&d_{0}&0\\
0&d_{1\mathbf{k}}&0&d_{0}
\end{pmatrix},
\end{eqnarray}
where $d_{0}=-0.1$ and $d_{1\mathbf{k}}=0.5\cos(k_{x}/2)\cos(k_{y}/2)$. However in Ref. \onlinecite{gaoy_fese} we did not consider the ellipticity of the $M$ electron pockets. Here we show, even in the presence of a slight ellipticity, the SC gap derived from our pairing function still agrees quite well with the experiment. As we can see from Fig. \ref{gap}(b), the gap magnitude is relatively larger on $\delta_2$ than that on $\delta_1$, while on both $\delta_1$ and $\delta_2$, the gap maxima (minima) are located along the $\Gamma-M$ ($X-M$) line. The gap anisotropy derived here is consistent with both the ARPES data \cite{shenzx2} and our previous result.\cite{gaoy_fese} In the inset of Fig. \ref{gap}(b) we show the density of states (DOS) in the SC state, where two pairs of the SC coherence peaks appear at $\pm\Delta_1$ (0.14, red arrows) and $\pm\Delta_2$ (0.17, blue arrows), respectively. This two gap structure is also consistent with STM experiments.\cite{xueqk1,fengdl3,xueqk4,xueqk5} Further inspection of the pairing function shows that, on the two electron bands, the SC order parameters are negative ($\Delta_{\mathbf{k}}^{3},\Delta_{\mathbf{k}}^{4}<0$) while on the hole bands, there exist sign changes in $\Delta_{\mathbf{k}}^{1}$ and $\Delta_{\mathbf{k}}^{2}$ (see Fig. \ref{gap_bz}). The pairing symmetry is generally $s$-wave and since the hole bands do not form Fermi surfaces, thus we denote this pairing symmetry as a hidden sign-changing $s$-wave symmetry. In Ref. \onlinecite{gaoy_fese} it was shown that this pairing symmetry can be distinguished from the conventional $s$-wave one by investigating the impurity-induced in-gap bound states. In this work we intend to study the possible spin excitation structure that may be observed by INS.

\begin{figure}
\includegraphics[width=1\linewidth]{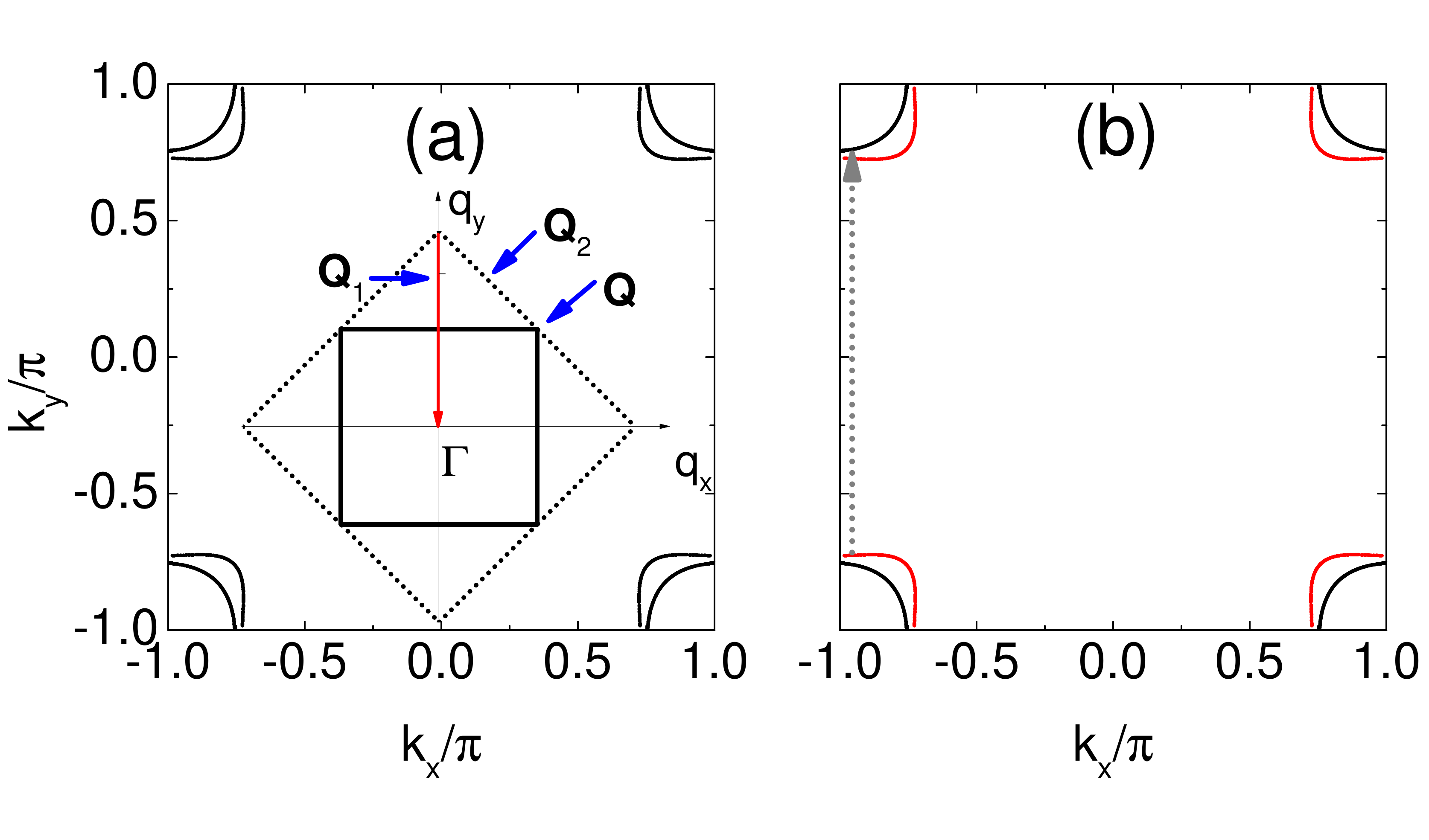}
 \caption{\label{sign} (color online) The sign of the SC order parameter on the Fermi surface. (a) is for cases \Rmnum{1} and \Rmnum{2}, while (b) is for case \Rmnum{3}. The black/red color denotes that the sign of the SC order parameter is negative/positive. In the inset of (a), the solid and dotted lines denote the 2Fe/cell BZ and 1Fe/cell BZ, respectively. In (b), the gray dotted arrow represents $\mathbf{q}=\mathbf{\mathbf{Q}_1}$ that connects the region of $\delta_1$ and $\delta_2$, where the SC order parameters have opposite signs.}
\end{figure}

In the following we consider three $s$-wave pairing cases. The pairing function of the first one is given by Eq. (\ref{dk}) (case \Rmnum{1}). For the second case, we artificially set $\Delta_{\mathbf{k}}^{1}$ and $\Delta_{\mathbf{k}}^{2}$ to be zero, i.e., we assume that the SC pairing does not take place in the two hole bands (case \Rmnum{2}) and this is the conclusion drawn in Ref. \onlinecite{bang2}. For the last case, in addition to setting $\Delta_{\mathbf{k}}^{1}$ and $\Delta_{\mathbf{k}}^{2}$ to be zero, we further set $\Delta_{\mathbf{k}}^{3}$ to be $-\Delta_{\mathbf{k}}^{3}$. In this way, there will exist a sign change of the SC order parameter between $\delta_1$ and $\delta_2$ (case \Rmnum{3}), as proposed in Ref. \onlinecite{chubukov} [see Fig. 4(g) of Ref. \onlinecite{chubukov}]. Considering the SC order parameters along the Fermi surfaces, in all the three cases, their magnitudes are consistent with ARPES and are the same as that shown in Fig. \ref{gap}(b) while their signs can be seen in Fig. \ref{sign}. Here we do not consider the nodeless $d$-wave pairing as proposed in Ref. \onlinecite{scalapino}, for two reasons. The first one is, as can be seen in Fig. 4 of Ref. \onlinecite{scalapino}, along one pocket, the gap magnitude of the nodeless $d$-wave pairing changes from the maxima to the minima as $\theta$ varies from $0$ to $\pi/2$, which is not consistent with the ARPES data since the gap magnitude should be both the local maxima at $\theta=0$ and $\pi/2$ [see Fig. 4(c) in Ref. \onlinecite{shenzx2}]. The inconsistency of the nodeless $d$-wave pairing can also be seen from Fig. 4(d) of Ref. \onlinecite{shenzx2}. Secondly, the nodeless $d$-wave pairing state requires that along $k_x=0$ and $k_y=0$ (in the 2Fe/cell BZ), $\Delta_\mathbf{k}$ should be zero. However in our proposed gap function, this cannot be satisfied as can be seen from Fig. \ref{gap_bz}. Therefore, we cannot assume any pairing function whose gap magnitude is consistent with ARPES while respects the nodeless $d$-wave pairing symmetry.

\begin{figure}
\includegraphics[width=1\linewidth]{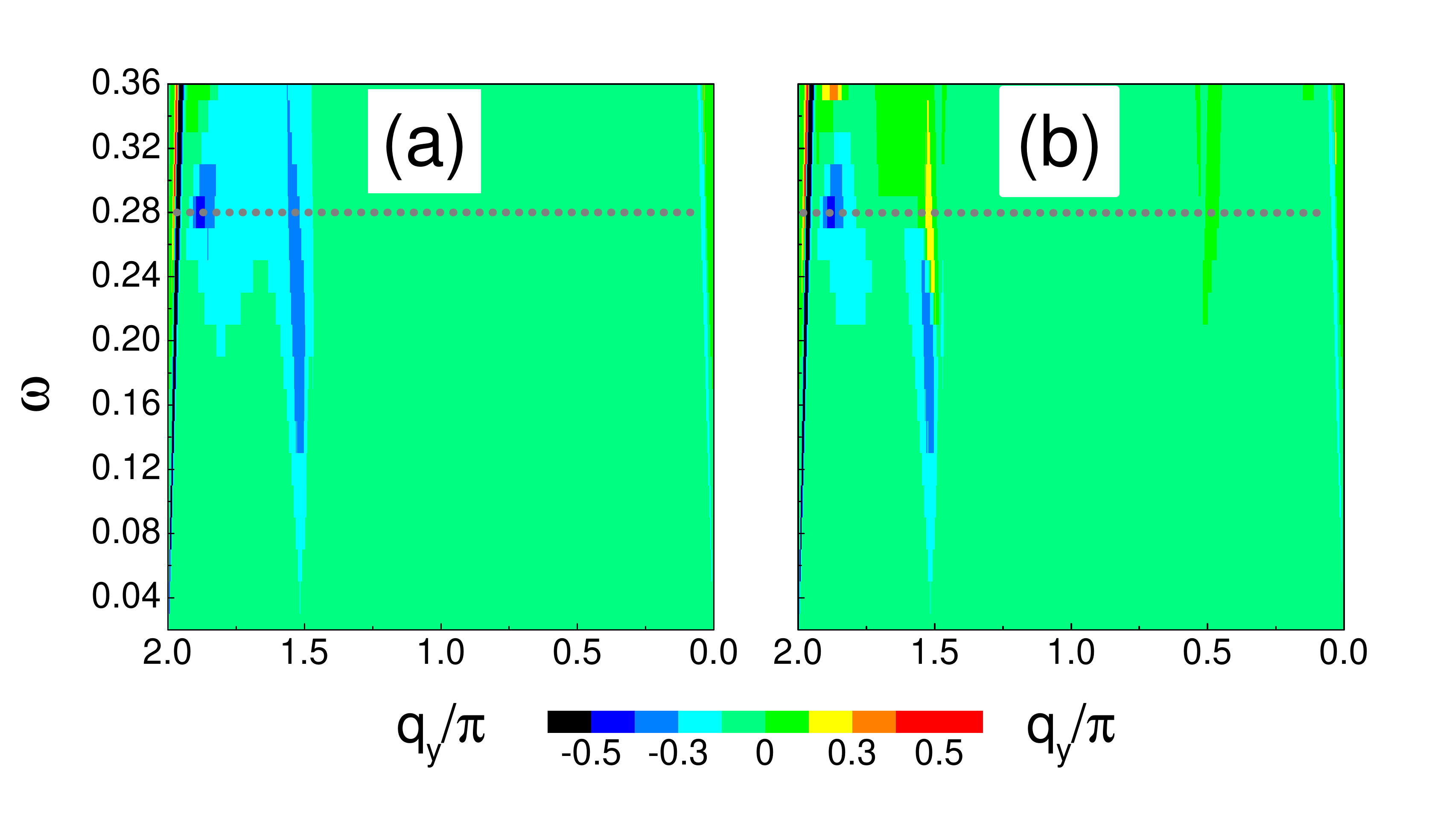}
 \caption{\label{alongqx=0} (color online) Difference of $Im\chi(\mathbf{q},\omega+i\eta)$ between the SC and normal states, calculated along the red arrow in the inset of Fig. \ref{sign}(a). The gray dotted lines denote $2\Delta_1=0.28$. (a) is for cases \Rmnum{1} and \Rmnum{2}, while (b) is for case \Rmnum{3}.}
\end{figure}

\begin{figure}
\includegraphics[width=1\linewidth]{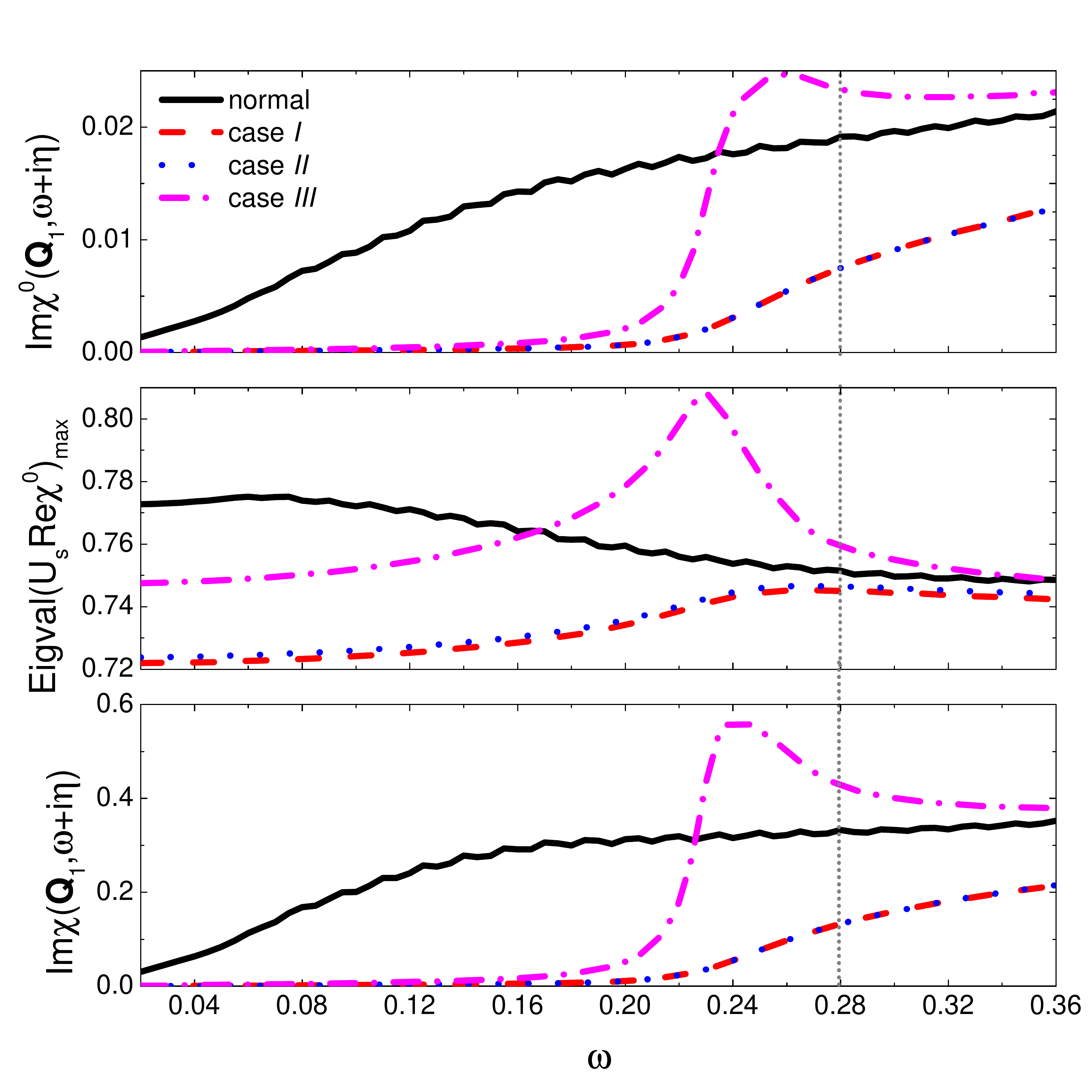}
 \caption{\label{q1} (color online) Spin susceptibility as a function of $\omega$, at $\mathbf{q}=\mathbf{Q}_1$. The gray dotted lines denote $\omega=2\Delta_1$.}
\end{figure}

In calculating $Im\chi(\mathbf{q},\omega+i\eta)$, we set $N=2048\times2048$ and $\eta=0.008$. In order to see the superconductivity-induced intensity change, we subtract $Im\chi(\mathbf{q},\omega+i\eta)$ in the SC state by its normal state value. We find that the intensity change is most prominent along the red arrow in the inset of Fig. \ref{sign}(a) and the results are shown in Fig. \ref{alongqx=0}. As we can see, for cases \Rmnum{1} and \Rmnum{2}, $Im\chi(\mathbf{q},\omega+i\eta)$ generally decreases once entering the SC state while for case \Rmnum{3}, there exists a clear increase of intensity close to $\mathbf{Q}_1\approx(0,1.505859375\pi)$ [the position of $\mathbf{Q}_1$ is shown in the inset of Fig. \ref{sign}(a)] and around $2\Delta_1$. To see it more clearly, we plot in Fig. \ref{q1} the spin susceptibility as a function of $\omega$, at $\mathbf{q}=\mathbf{\mathbf{Q}_1}$. We find that the spin susceptibilities in the normal state, as well as in cases \Rmnum{1} and \Rmnum{2}, are mostly featureless. In contrast, for case \Rmnum{3}, there exists a step-like jump in $Im\chi^0$ around $\omega\approx0.23$, which leads to a peak in the maximal eigenvalue of $U_sRe\chi^0$ at the same $\omega$. The origin of these is because, at $\mathbf{q}=\mathbf{\mathbf{Q}_1}$, the threshold energy $E_{th}$ due to the $\delta$ function in Eq. (\ref{imx0}) is approximately determined by $|\Delta_{\mathbf{k}}^{3}|+|\Delta_{\mathbf{k}+\mathbf{Q}_1}^{4}|$, where $\mathbf{k}$ and $\mathbf{k}+\mathbf{Q}_1$ are momenta on $\delta_2$ and $\delta_1$, respectively, as denoted by the gray arrow in Fig. \ref{sign}(b). In both cases \Rmnum{1} and \Rmnum{2}, the coherence factor $C^{34}(\mathbf{k},\mathbf{Q}_1)$ associated with this process, is zero since $E_{\mathbf{k}}^{3},E_{\mathbf{k}+\mathbf{Q}_1}^{4}=0$ and $sgn(\Delta_{\mathbf{k}}^{3})=sgn(\Delta_{\mathbf{k}+\mathbf{Q}_1}^{4})$. On the contrary, in case \Rmnum{3}, since $sgn(\Delta_{\mathbf{k}}^{3})=-sgn(\Delta_{\mathbf{k}+\mathbf{Q}_1}^{4})$, the above mentioned coherence factor $C^{34}(\mathbf{k},\mathbf{Q}_1)$ reaches its maximal value two. Therefore in cases \Rmnum{1} and \Rmnum{2}, the step-like jump in $Im\chi^0$, as well as the peak in the maximal eigenvalue of $U_sRe\chi^0$, are suppressed while they are enhanced in case \Rmnum{3}. Careful inspection of $Im\chi$ shows that, in cases \Rmnum{1} and \Rmnum{2}, $Im\chi$ decreases once entering the SC state and no resonance-like structure exists at $\omega\leq2\Delta_1$. In comparison, for case \Rmnum{3}, a broad hump centered around $\omega\approx0.245$ shows up, whose intensity exceeds its normal-state counterpart while at $\omega\leq0.23$, $Im\chi$ is suppressed in the SC state compared to that in the normal state. Since the hump structure is located at $\omega<2\Delta_1$, at a first glance, it seems that a spin resonance exists in case \Rmnum{3}. However as we can see, the hump around $\omega\approx0.245$ in $Im\chi$ already exists in $Im\chi^0$, whose maximum is located at $\omega\approx0.26$ while the profiles of $Im\chi^0$ and $Im\chi$ are similar, except for an increased intensity in $Im\chi$. The hump structure in $Im\chi$ therefore does not signify a spin resonance since at $\omega\approx0.245$, $Im\chi^0$ is not close to zero. Furthermore, since the largest maximal eigenvalue of $\hat{U}_sRe\hat{\chi}^{0}(\mathbf{Q}_1,\omega+i\eta)$ is $\sim0.81$ at $\omega=0.23$, the condition $det[\hat{I}-\hat{U}_sRe\hat{\chi}^0]=0$ cannot be met for any $\omega$. Therefore the collective spin excitation at $\omega\approx0.245$ and $\mathbf{q}=\mathbf{Q}_{1}$ is highly damped. In INS experiments, when entering the SC state, an intensity increase at $\omega<2\Delta$ is commonly attributed to the formation of spin resonances, where usually $\Delta$ is the STM-measured SC gap. Here we show, since the SC order parameters are anisotropic on the Fermi surfaces, $Im\chi^0$ can be nonzero at $\omega\leq2\Delta_1$ while it will be zero if the SC gaps are isotropic. Consequently, at $\omega<2\Delta_1$, there may exist an increase of $Im\chi$ when entering the SC state as long as $Im\chi^0$ is increased compared to its normal-state value and this intensity increase does not signify a spin resonance. As shown for case \Rmnum{3}, the hump structure in $Im\chi$ is very robust since it already appears in $Im\chi^0$. Even if the exact value of $U$ is unknown, for $0\leq U<6.8$, the hump structure will always appear at $\omega<2\Delta_1$. Therefore, cases \Rmnum{1} and \Rmnum{2} can be distinguished from case \Rmnum{3} by their different behaviors at $\mathbf{Q}_1$.

\begin{figure}
\includegraphics[width=1\linewidth]{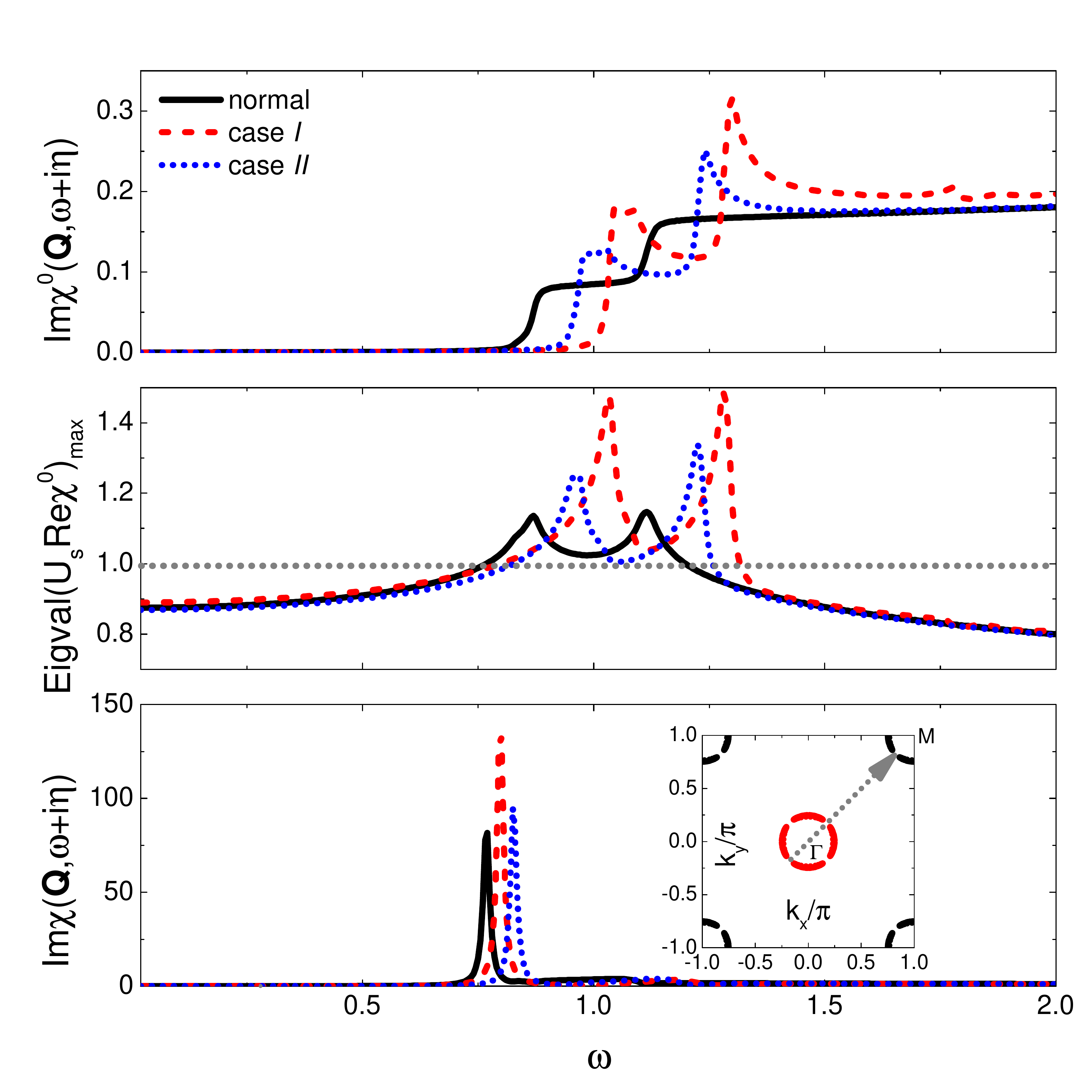}
 \caption{\label{q} (color online) Spin susceptibility as a function of $\omega$, at $\mathbf{q}=\mathbf{Q}$. The inset shows schematically the scattering process connected by $\mathbf{Q}$ (gray dotted arrow). Here the red pocket around $\Gamma$ is incipient, which sinks below the Fermi level.}
\end{figure}

In order to investigate whether cases \Rmnum{1} and \Rmnum{2} can be distinguished from each other by INS, we study the spin susceptibilities at $\mathbf{Q}$ [the position of $\mathbf{Q}$ can be seen in the inset of Fig. \ref{sign}(a)]. As we can see from the inset of Fig. \ref{q}, this wave vector connects the momenta on the $M$ electron pockets with those on the $\Gamma$ hole bands which sink below the Fermi level. In the usual iron pnictide superconductors, INS experiments commonly found a spin resonance at this wave vector, which supports the sign-changing characteristics of the SC order parameter between the $M$ electron and $\Gamma$ hole pockets, therefore naively we expect the same phenomenon may occur in case \Rmnum{1}, but not case \Rmnum{2}. Numerical results in Fig. \ref{q} shows that, already in the normal state, there exist two step-like jumps in $Im\chi^0$ and two peaks in the maximal eigenvalue of $\hat{U}_sRe\hat{\chi}^0$, which are related by the KK relations. Below the first step, $Im\chi^0$ is zero since the hole bands sink below the Fermi level and the threshold energy associated with the scattering process in the inset of Fig. \ref{q} must overcome the energy gap between the top of the hole band (band 2) and the Fermi level. Similarly, the second step jump in $Im\chi^0$ is resulted from the scattering between the $M$ electron pockets and another hole band (band 1) which is even deeper below the Fermi level. Similar structures exist in cases \Rmnum{1} and \Rmnum{2} as well, which seems to be insensitive  to the relative sign of the SC order parameters connected by $\mathbf{Q}$. The reason is that, suppose $\mathbf{k}$ is on the $M$ electron pockets and $\mathbf{k}+\mathbf{Q}$ is on the $\Gamma$ hole bands, then in the coherence factor we have $E_{\mathbf{k}}^{o}\approx0$ and $\xi_{\mathbf{k}}^{o}\approx|\Delta_{\mathbf{k}}^{o}|$, leading to $C^{op}(\mathbf{k},\mathbf{Q})\approx1+\frac{\Delta_{\mathbf{k}+\mathbf{Q}}^{p}}{\xi_{\mathbf{k}+\mathbf{Q}}^{p}}$. Meanwhile since the hole bands are located at $\sim100$meV below the Fermi level, we further have $|\Delta_{\mathbf{k}+\mathbf{Q}}^{p}|\ll|E_{\mathbf{k}+\mathbf{Q}}^{p}|$. Therefore in this case $C^{op}(\mathbf{k},\mathbf{Q})$ is in the vicinity of one, irrespective of the sign of $\Delta_{\mathbf{k}+\mathbf{Q}}^{p}$ and the step jumps in $Im\chi^0$, as well as the peaks in the maximal eigenvalue of $\hat{U}_sRe\hat{\chi}^0$, will also exist in cases \Rmnum{1} and \Rmnum{2}. Similar behaviors can also be found in $Im\chi$, where a peak shows up in the normal state, as well as in cases \Rmnum{1} and \Rmnum{2}. Here the peaks are truly the signatures of spin resonances since the conditions $Im\chi^0=0$ and $det[\hat{I}-\hat{U}_sRe\hat{\chi}^0]=0$ can be simultaneously met. For $U=6$ as we considered, the crossings between the gray dotted line and the curves of the maximal eigenvalue of $\hat{U}_sRe\hat{\chi}^0$ in the middle panel of Fig. \ref{q} determine the location of the peaks in $Im\chi$. As $U$ decreases, the gray dotted line will shift up and the peaks in $Im\chi$ will move to higher energy. Finally if $U$ is infinitesimally small, $Im\chi$ will be the same as $Im\chi^0$. Therefore, at $\mathbf{q}=\mathbf{Q}$, in contrast to our naive expectation, the spin susceptibilities in cases \Rmnum{1} and \Rmnum{2} are similar and cannot be distinguished from each other.

Finally, we comment on the spin susceptibilities at $\mathbf{q}=\mathbf{Q}_2$, whose location is shown in the inset of Fig. \ref{sign}. A spin resonance at this wave vector has been theoretically predicted \cite{scalapino} and experimentally observed in Li$_{1-x}$Fe$_x$ODFe$_{1-y}$Se,\cite{Boothroyd,zhaoj1} Li$_x$(ND$_2$)$_y$(ND$_3$)$_{1-y}$Fe$_2$Se$_2$ \cite{Boothroyd1} and A$_x$Fe$_{2-y}$Se$_2$ (A=Rb, Cs, K).\cite{Inosov1,Inosov2,Boothroyd2,Inosov3,zhaoj} Beside the coherence factor, the existence of this spin resonance relies closely on the square or rectangular shape of the Fermi surfaces. As shown in Fig. 4(b) of Ref. \onlinecite{Inosov2} and Fig. 7 of Ref. \onlinecite{Boothroyd}, the scatterings between the flat parts of the Fermi surfaces lead to the nesting wave vector $\mathbf{Q}_2$. However, as we can see from Fig. 7 of Ref. \onlinecite{Boothroyd}, the Fermi surfaces are perpendicular to the $\Gamma-X$ line, i.e., if viewing in the 2Fe/cell BZ, $\delta_1$ and $\delta_2$ are both perpendicular to the $\Gamma-M$ line, as shown by the blue dotted lines in Fig. \ref{band}(b). Apparently this is not the case in monolayer FeSe grown on SrTiO$_{3}$ since high-resolution ARPES found elliptical Fermi surfaces with no flat parts. \cite{shenzx2} Indeed, the calculated spin susceptibilities at $\mathbf{Q}_2$ are negligibly small compared to those at $\mathbf{Q}_1$ (not shown here), suggesting that there will not exist spin resonances at $\mathbf{Q}_2$ in monolayer FeSe grown on SrTiO$_{3}$. In contrast, the spin excitation should be peaked at $\mathbf{Q}_1$, which connects the parts of the electron pockets close to the BZ boundaries, similar to that shown in Fig. 1(a) of Ref. \onlinecite{leedh2}. In addition we infer, even if the pairing symmetry is a nodeless $d$-wave, as schematically shown in Fig. 4(h) of Ref. \onlinecite{chubukov}, $Im\chi$ should still be peaked around $\mathbf{Q}_1$, similar to case \Rmnum{3}.

\section{summary}
In summary, we have investigated possible spin excitation spectra in monolayer FeSe grown on SrTiO$_{3}$, based on a model fitting the ARPES-measured band structure and gap anisotropy. We found that, if the SC order parameter changes sign between the inner and outer pockets, a resonance-like spin excitation may occur, which is highly damped. On the contrary, if the SC order parameter maintains its sign along the Fermi surfaces, no such spin excitation can occur. Furthermore, since the hole bands sink $\sim$100meV below the Fermi level, INS experiments are unable to tell whether or not Cooper pairs form on the hole bands, let alone the sign of the SC order parameter. Finally we propose that the observed spin resonance at $\mathbf{Q}_2$ in other FeSe-based superconductors may not exist in monolayer FeSe grown on SrTiO$_{3}$, since the Fermi surfaces in this material are elliptical, rather than square or rectangular-like. Possibly the SC pairing symmetry in monolayer FeSe grown on SrTiO$_{3}$ may also be different from other FeSe-based superconductors, since at least in Li$_{1-x}$Fe$_x$OHFeSe, the maxima and minima of the gap are located at exactly the opposite locations compared to monolayer FeSe grown on SrTiO$_{3}$ (see Fig. 7 of Ref. \onlinecite{wenhh}). Beside INS, the spin excitation spectra proposed in this work may also be measured by STM in real space as proposed in a recent work.\cite{Wahl}

\section*{ACKNOWLEGEMENTS}
This work is supported by the Natural Science Foundation from Jiangsu
Province of China (Grants No. BK20160094 and No. BK20141441), the NSFC (Grant No. 11374005) and NSF of Shanghai (Grant No. 13ZR1415400). QHW is supported by NSFC (under grant No.11574134).

\end{document}